\documentclass[preprint,showpacs,preprintnumbers,amsmath,amssymb]{revtex4}

\usepackage{graphicx}
\usepackage{dcolumn}
\usepackage{bm}

\begin{document}

\preprint{}

\title{
Voltage-Selective Bi-directional Polarization and Coherent Rotation of Nuclear Spins in Quantum Dots}

\author{R. Takahashi$^{1,2}$, K. Kono$^{1,2}$, S. Tarucha$^{3}$, and K. Ono\footnote{ E-mail address: k-ono@riken.jp}$^{1,4}$ }
\affiliation{$^1$Low Temperature Physics Laboratory, RIKEN, Wako, Saitama 351-0198, Japan\\
$^2$Department of Physics, Tokyo Institute of Technology, Meguro, Tokyo 152-8551, Japan\\
$^3$Department of Applied Physics, University of Tokyo, Bunkyo, Tokyo 113-8656, Japan}

\date{\today}

\begin{abstract}

We proposed and demonstrated that the nuclear spins of the host lattice in GaAs double quantum dots can be strongly polarized in either of two opposite directions, parallel or antiparallel to an external magnetic field. The direction is selected simply by adjusting the dc source-drain voltage of the device. This nuclear polarization manifests itself by repeated controlled electron-nuclear spin scattering in the Pauli spin blockade state. Polarized nuclei are also controlled coherently by means of nuclear magnetic resonance (NMR). This work confirmed that the nuclear spins in quantum dots are indeed long-lived quantum states with a coherence time of up to 1 ms, and may be a promising resource for quantum information processing such as quantum memories for electron spin qubits.

\end{abstract}

\maketitle

Nuclear spins are known to have a long coherence time and are a promising resource for quantum information processing by standard and optically detected NMR \cite{awschalom}. Recently, nuclear spins in semiconductor devices have attracted much attention because the quantum resource is electrically accessible \cite{kronmuller1, machida, yusa, simmons}. The detection and coherent control of nuclear spins have been implemented in two-dimensional electron devices using the quantum Hall regimes. In these devices, ensembles of spatially delocalized electrons are used to generate and detect nuclear polarization, which provides limited controllability of nuclear spins in space. In quantum dot devices, on the other hand, nuclear spin dynamics are caused by an interaction with an electron with a well-defined orbital and spin state and provide a more controllable system \cite{xu, hanson}.

In previous works using double quantum dots (DQDs), nuclear spins were treated as a fluctuating effective magnetic field that affects the coherence of electron spin \cite{koppens, petta1, reilly1}. Recently, it has been reported that the coherence time of electron spin qubits is improved by treating the nuclear spin background \cite{reilly1}. However, direct control of the quantum nature of nuclear spins in a quantum dot has not yet been achieved, although such control has a great potential for nuclear spin bath control and increasing the coherence time of a coupled spin system \cite{DasSarma, hanson2}.

The rapid buildup of large polarization, up to 40\% within 1 s, of nuclear spin in a quantum dot has been achieved by electrical pumping in vertical DQDs \cite{baugh}. Furthermore, the realization of 100\% polarization is predicted using DQDs with realistic device parameters \cite{baugh}. Quantum memory for electron spin qubits has been theoretically investigated using  fully polarized nuclear spins in quantum dots \cite{taylor}. However, this large nuclear polarization in vertical DQDs has so far been fixed to only one direction, antiparallel to the magnetic field \cite{ono2, baugh}. Similar electrical pumping of nuclear spin has been attempted with lateral DQDs using a pulsed gate voltage sequence, but the polarization was limited to a small value of $\sim$ 1\%, which is far below the full polarization required by nuclear spin memory \cite{petta2, foletti}. So far, the \textit{difference} in the nuclear polarization vector between the two dots has been controlled bidirectionally within the range of small polarization of $\sim$ 1\% \cite{foletti}. This difference can be explained in terms of the diffusion time of the nuclear polarization in vertical ($\sim 200$ s) and lateral ($\sim 10$ s) DQDs, and the cycle rate of electron spin-nuclear spin scattering \cite{reilly1, takahashi1}.

\begin{figure}[t]
\begin{center}
\includegraphics[width=0.7 \textwidth, keepaspectratio]{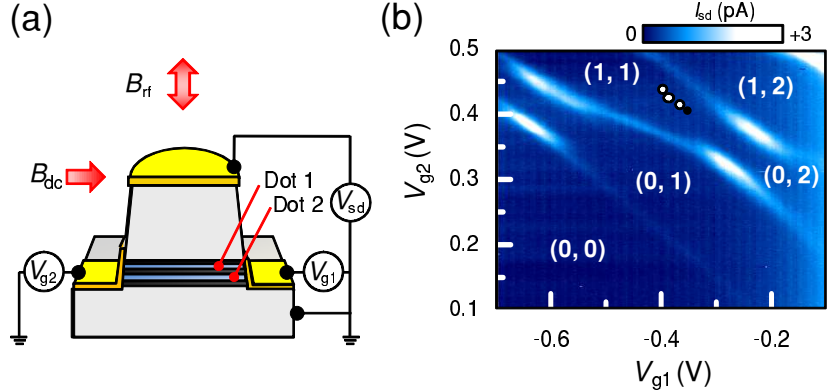}
\caption{(a) Schematic of our vertical DQD device. (b), Intensity plot of $I_{sd}$ measured with $V_{sd}=$ 0.1 mV, showing charge stability diagram.}
\label{device}
\end{center}
\end{figure}

In this letter, we report that nuclear spins in a vertical DQD can be strongly polarized either parallel or antiparallel to the magnetic field. The direction of nuclear polarization can be selected simply by tuning the dc source-drain voltage, $V_{sd}$, in the mV range. We also show that the nuclear spins are coherently rotated with a pulsed radio-frequency (RF) magnetic field, and that the rotation can also be detected electrically by means of the source-drain current, $I_{sd}$, of the device. The coherence time of nuclear spins in solids reaches the order of ms. Combining the $\pi$-pulse of this electrically detected NMR, we unambiguously show that the two directions of polarization are indeed opposite each other.

Dynamic nuclear polarization in quantum dots is accompanied by the relaxation of the Pauli spin-blockade (SB) state \cite{ono1, koppens}. The electron transport is blocked when two-electron spin triplets occupy the double dot, and the small leakage current in the SB state is a sensitive probe for spin scattering from the triplet. Electron spin-flips can be accompanied by flops of nuclear spins in the dot via a hyperfine interaction. By repeating these flip-flop processes, the nuclear spins of the dot are eventually polarized \cite{ono2, baugh}. The direction of polarization depends on the $z$-component of the initial triplet spin state, $S_z$. If $S_z$ can be selected by adjusting the dc voltage, the direction of nuclear polarization can be electrically controlled. In our experiment, a change in the nuclear spin state in DQDs was detected using a steplike increase of the leakage current in the SB regions \cite{ono2, baugh}.

Figure \ref{device}(a) shows a schematic of a vertical DQD device. The device is made of a sub-micron pillar structure that consists of an In$_{0.04}$Ga$_{0.96}$As quantum well, a GaAs quantum well and three Al$_{0.3}$Ga$_{0.7}$As barriers. This device has two Au/Ti Schottky gate electrodes biased with gate voltages $V_{g1}$ and $V_{g2}$. Quantum dots are formed in the quantum wells at the center of the pillar owing to the gate voltage confinement. The two gate electrodes have different capacitive couplings to the two dots because of their asymmetric shapes. Thus, the electrostatic potential of each dot can be tuned independently. Measurements are performed at $\sim$ 1.6 K in a pumped $^4$He cryostat. A static magnetic field, $B_{dc}$, is applied in an in-plane direction, and an ac magnetic field with radio frequency, $B_{rf}$, is applied perpendicular to the quantum wells with a coil placed  $\sim$ 1 mm above the device \cite{takahashi1}. Figure \ref{device}(b) shows an intensity plot of $I_{sd}$ as a function of $V_{g1}$ and $V_{g2}$. The plot exhibits the well-known honeycomb structure typical of weakly coupled quantum dots \cite{hanson}. This enables us to assign the ground-state charge configuration, $(n, m)$, of the two dots, where $n$ and $m$ are the numbers of electrons in dots 1 and 2, respectively. The SB appears in the region defined by the points (1, 1), (0, 2) and (0, 1) under a finite $V_{sd}$.

\begin{figure}[t!]
\begin{center}
\includegraphics[width=0.6 \textwidth, keepaspectratio]{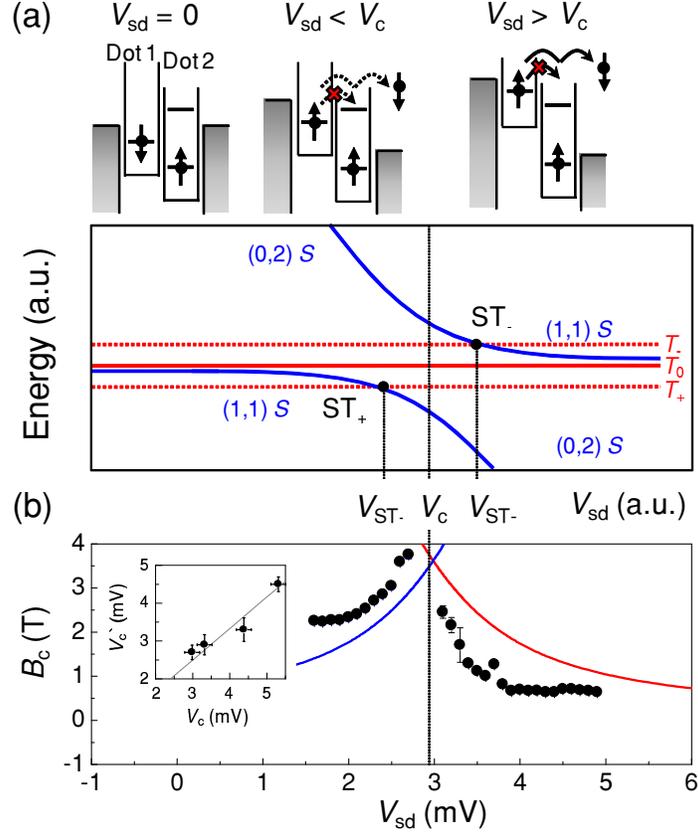}
\caption{(a) Two-electron energy diagram at the position of the filled circle in Fig. \ref{device}(b). Blue and red lines show the singlet and Zeeman-split triplet states, respectively. Upper schematics of energy landscapes of the device at $V_{sd} = 0$ (left), thermally assisted SB (middle) and ordinary SB (right). Gray arrows describe allowed and forbidden tunneling processes for singlet or triplet spins. (b) Plot of magnetic field where a steplike increase of the leakage current is observed. Solid curves are crossover points between the singlet and triplet calculated with the device parameters given in the caption of Fig. \ref{device}(b). (Left inset) $V_c$ evaluated from Fig. \ref{device}(b) and $V'_c$ measured by measurements similar to those in the case of Fig. \ref{diagrams}(b).}
\label{diagrams}
\end{center}
\end{figure}

Figure \ref{diagrams}(a) shows a schematic of the two-site, two-electron energy diagram that corresponds to the filled circle in Fig. \ref{device}(b). The SB eventually fixes the spin state to either $T_-$, $T_0$ or $T_+$ with a charge configuration of (1, 1). The horizontal axis in Fig. \ref{diagrams}(a) is the energy difference between dots, which can be linearly tuned by $V_{sd}$. The ground state is the spin singlet with a charge configuration of (1, 1), \textit{i.e.}, (1, 1)$S$, for $V_{sd} \sim 0$ , or a singlet with (0, 2), \textit{i.e.}, (0, 2)$S$, for large $V_{sd}$. The value of $V_{sd}$ for an anticrossing point with two singlets, $V_c$, can be estimated from Fig. \ref{device}(b) and Coulomb diamond measurements \cite{takahashi1}, and it is  3.0 ($\pm$ 0.3) mV. Two types of singlet-triplet crossover points, denoted as $ST_-$ and $ST_+$, are achieved at $V_{sd} = V_{ST-}$ and $V_{sd} = V_{ST+}$, respectively, for a given $B_{dc}$. An ``ordinary" SB appears for $V_{sd} > V_c$. The region between $\sim$ 1.7 mV and $V_c$ corresponds to a ``thermally assisted" SB, which is essentially the same as the ordinary SB except that the interdot tunneling from (1, 1)$S$ to (0, 2)$S$ is accompanied by the absorption of phonons (see inset).

In the ordinary SB region, where $V_{sd} > V_c$, upon sweeping $B_{dc}$ upward from 0 T, the leakage current is known to increase in a stepwise manner at a certain magnetic field, $B_c$ \cite{ono2, baugh}. At $B_c$, the energy differences between (1, 1)$S$ and $T_-$ are compensated by Zeeman splitting of the triplets, leading to elastic scattering from $T_-$ to (1, 1)$S$ accompanied by scattering of a nuclear spin. In this region,  $B_c$ \textit{decreases} with increasing $V_{sd}$, because the energy difference between (1, 1)$S$ and $T_-$ decreases \cite{ono2, baugh}.

The steplike increase in the leakage current was also observed for  $V_{sd} < V_c$.  The measured $B_c$ \textit{increases} with $V_{dc}$ in this region, and results in a peak of the plot at $V_{sd} = V'_c\sim 2.8$ mV, as shown in Fig. \ref{diagrams}(b).  The steplike change of $I_{sd}$ for  $V_{sd} < V_c$ is reasonably explained as elastic scattering from $T_-$ to (0, 2)$S$.  Solid curves in the main panel of Fig. \ref{diagrams}(b) show $B_c$ calculated using the same device parameters as those for the $T_+$ - (0, 2)$S$ crossing (blue line) and the $T_-$ - (1, 1)$S$ crossing (red line). They are in reasonable agreement with the measured values of $B_c$. We implement similar measurements as in the case of Fig. \ref{diagrams}(b) at different gate voltages, indicated by the open circles in Fig. \ref{device}(b), and compare the obtained values of $V'_c$ with the calculated values. $V_c$ and $V'_c$ are nearly identical, as shown in the left inset of Fig. \ref{diagrams}(b). 

If the electron state is (1, 1)$S$, it relaxes rapidly to (0, 2)$S$ and sequentially changes to the (0, 1) state with a typical time scale of $\sim$ 1 ns in our device. Then, the next electron tunnels from the source electrode to the left dot, and a new (1, 1) state is loaded. On the other hand, the triplet states relaxes directly to the (0, 1) state via a cotunneling process with a typical time interval of $\sim$ 100 ns, and again, the next (1, 1) state is loaded. At $B_c$, the scattering from one triplet to  (1, 1)$S$ is allowed with the spin flip-flop of electron and nuclear spins via a hyperfine interaction. For a given magnetic field, at $V_{ST_-}$, the change in the electron spin angular momentum is $+1$, thus that in the nuclear spin is $-1$. Consequently, if the flip-flop process from $T_-$ to (1, 1)$S$ is repeated, nuclear spins polarize ``downward". Note that the occupation of $T_0$ or $T_+$ does not contribute to this nuclear pumping, and these unnecessary triplets are refreshed by the cotunneling process. Thus, the total magnetic field increases with polarizing nuclear spins ($B_n$, nuclear field) to $B_{dc} +B_n$. Because there is the relation $A\langle I_z\rangle =g\mu_B B_n$ and the negative sign of the electron g-factor for the (In)GaAs dot, where $A$ is a hyperfine coupling constant and $I_z$ is the z-component of a nuclear spin. We call this a parallel polarization. Similarly, at $V_{ST_+}$, an antiparallel polarization is induced owing to a repetition of the flip-flop from $T_+$ to (1, 1)$S$ and the cotunneling-induced refreshment of $T_0$ or $T_-$. Thus, nuclear spins can be polarized either parallel or antiparallel to the external field, and the direction is switched simply by changing $V_{sd}$.

After the steplike leakage current increase at $B_c$, the system shows complicated behavior such as hysteresis with a downward sweep of $B_{dc}$ or long-term oscillation \cite{ono2, baugh}. These behaviors may be due to nontrivial feedback from polarized nuclear spins, which is not explained by a simple Overhauser model. In this letter, we focus only on the region where the stepwise increase of the leakage current is observed, where the leakage current takes only two steady-state values, \textit{i.e.}, before and after the step increase, and we regard these two currents as those before and after polarization, respectively.

\begin{figure}[t]
\begin{center}
\includegraphics[width=0.7 \textwidth, keepaspectratio]{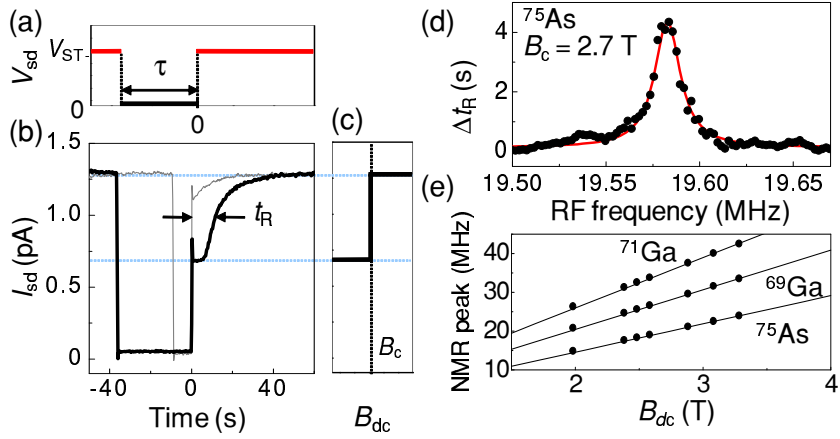}
\caption{(a) Schematic of $V_{sd}$ switching between $V_{ST_-}$ and 0 ($ST_- -ST_-$ sequences). (b) Transient behavior of the leakage current in $ST_- -ST_-$ sequences for $V_{ST_-}=$ 3.6 mV with $\tau = 37$ s (solid line) and $\tau = 5$ s (gray line). (c) Schematic of the current for up-sweeping $B_{dc}$, showing the step at $B_c$.   (d) Plot of the change in $t_R$, $\Delta t_R$. Each point was obtained from the sequences with $\tau =$ 3 s in continuous wave $B_{rf}$ with various frequencies at $B_{dc} =$ 2.7 T. (e) $B_{dc}$ dependence of $t_R$ peaks that correspond to nuclear species $^{71}$Ga, $^{69}$Ga and $^{75}$As.}
\label{nmr}
\end{center}
\end{figure}

Under a finite magnetic field, suppose that $V_{sd}$ is switched from $V_{ST_-}$ to 0, and after a dwell time, $\tau$, is switched back to $V_{ST_-}$. By this operation  (a $ST_- -ST_-$ sequence, see Fig. \ref{nmr}(a)), we found that $I_{sd}$ exhibits transient behavior \cite{takahashi1, baugh}. The solid line in Fig. \ref{nmr}(b) shows the behavior of $I_{sd}$ when $V_{sd}$ is switched from $V_{ST_-} =$ 3.6 mV to 0 mV at time $-37$ s, then switched back to $V_{ST_-}$ when the time is 0 s. In this case, the current is held at the lower value ($I_{sd} \sim$ 0.7 pA) for the first several seconds and then is shifted to a higher value ($I_{sd} \sim$ 1.3 pA) after a characteristic recovery time, $t_R$. We found that the lower and higher current values correspond to the values before and after the current step at $B_c$, respectively, as shown in the schematic magnetic field dependence (Fig. \ref{nmr}(c)). On the other hand, the current switches back to the higher value immediately ($t_R \sim$ 0 s) for short $\tau$, as shown by the gray line in Fig. \ref{nmr}(b). We assign $t_R$ to the peak of the derivative with respect to time of the leakage current, $dI_{sd}/dt$. We measured $t_R$ for  $ST_- -ST_-$ sequences with various $\tau$, and found that $t_R$ increases linearly with increasing $\tau$ and eventually saturates for long $\tau$. The behavior can be fitted by exponential decay with a characteristic decay time $T_D$ of $\sim$ 200 s \cite{takahashi1}. In Fig. \ref{nmr}(d), we repeat a similar sequence with various frequencies in a continuous wave (CW) $B_{rf}$ for $\tau \ll T_D$. We found an enhancement of $t_R$ at approximately 19.57 MHz, which is the NMR frequency of $^{75}$As. Similar peaks were detected at NMR frequencies of both $^{69}$Ga and $^{71}$Ga. These peaks indeed shift linearly with $B_{dc}$, as shown in Fig. \ref{nmr}(e). Similar behaviors to those shown in Fig. \ref{nmr}(d) and e are also observed in an $ST_+ -ST_+$ sequence, \textit{i.e.}, $V_{sd}$ is switched from $V_{ST_+}$ to 0 and back to $V_{ST_+}$.

On the basis of the above-described behaviors, $t_R$ can be interpreted as the typical time required to return to the initial polarized nuclear spin state, and is in proportional to the degree of depolarization of nuclear spins. In  $ST_- -ST_-$ ($ST_+ -ST_+$) sequences, nuclear spin polarization is saturated at $V_{ST_-}$ (antipolarized or negatively polarized at $V_{ST_+}$), and then nuclear spin pumping (antipumping) is stopped at $V_{sd} =$ 0 mV. In Fig. \ref{nmr}(b), nuclear spins remain polarized for small values of $\tau$, start to depolarize with increasing $\tau$ and are fully depolarized for $\tau \gg T_D$. The long decay time on the order of minutes is consistent with those observed in quantum Hall systems \cite{kronmuller1}. Note that $V_{sd}$ is initially held at $V_{ST_-}$ (or $V_{ST_+}$) for a sufficiently long duration ($\sim$ 180 s) to ensure well-defined initial states and that the series of $t_R$ measurements is not affected by the previous measurement. It is known that polarized nuclear spins are eventually depolarized if $B_{rf}$ with NMR frequency is applied continuously. In Fig. \ref{nmr}(d), for off-resonant frequencies, nuclear spins retain their polarization because of the sufficiently short $\tau$. However, at the NMR frequency, nuclear spins rapidly depolarize, leading to the enhancement of $t_R$. Although the result in Fig. \ref{nmr}(d) can be interpreted on the basis of the polarized (antipolarized) nuclear spins at $V_{ST_-}$ ($V_{ST_+}$), there is no direct evidence that these polarized states are indeed antiparallel to each other; to be exact, measurements such as those in Fig. \ref{nmr} only indicate that some nuclear spin orders appeared at $V_{ST_-}$ and $V_{ST_+}$, and that these orders can be destroyed for $\tau \gg T_D$ or a CW $B_{rf}$ with NMR frequencies.

\begin{figure}[t]
\begin{center}
\includegraphics[width=0.8\textwidth, keepaspectratio]{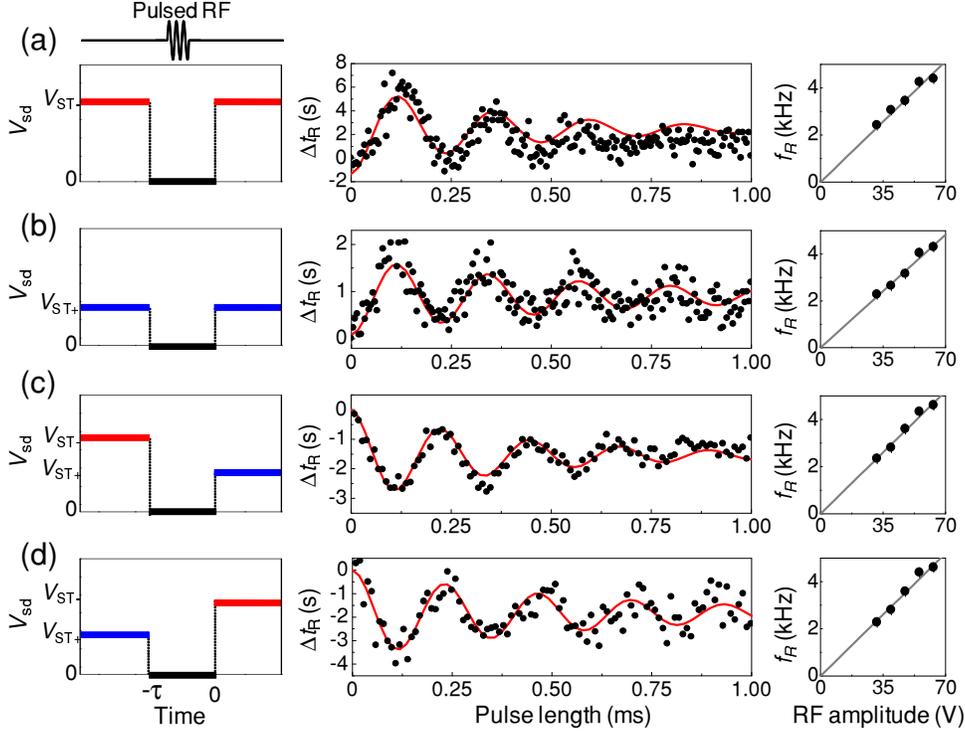}
\caption{$V_{sd}$ sequences (left panels), pulse length dependence of $\Delta t_R$ for the peak-to-peak amplitude of $B_{rf}$ at 55 V (middle panels), and the amplitude dependence of Rabi frequency, $f_R$ (right panels), for (a) $ST_- -ST_-$, (b) $ST_+ -ST_+$, (c) $ST_- -ST_+$ and (d) $ST_+ -ST_-$ sequences.  For all sequences, $B_{dc} =$ 2.6 T, where $V_{ST_-}$ and $V_{ST_+}$ are 3.2 mV and 2.1 mV, respectively.  A $B_{rf}$ with 18.5 MHz, which is the NMR frequency of $^{75}$As, is used. $\tau$ is 30 s for (a) and 3 s for (b) - (d). All plots are fitted by a sinusoidal curve with exponential decay (red curves).}
\label{rabi}
\end{center}
\end{figure}

To confirm the above points, we employed a pulse-modulated $B_{rf}$. It is known that nuclear spins are rotated coherently if a strong pulse-modulated $B_{rf}$ with NMR frequency is applied.  The nuclear spins are reversed if the so-called $\pi$-pulse is applied. Hereafter, we show that the nuclear spin order at $V_{ST_-}$ indeed equals the nuclear spin order at $V_{ST_+}$ \textit{with the} $\pi$-pulse, and vice versa. 

In a series of  $ST_- -ST_-$ sequences used for pulsed $B_{rf}$ measurement, we apply a pulse-modulated $B_{rf}$ with various pulse lengths of 0 - 2 ms at the time of $\sim -$1.5 s at the stage when $V_{sd} = 0$ mV (Fig. \ref{rabi}(a) left panel). The three steps of the $ST_- -ST_-$ sequence correspond to the initialization, operation and detection of nuclear spins. The measured $t_R$ exhibits a sinusoidal (Rabi) oscillation as a function of the pulse length (Fig. \ref{rabi}(a) middle panel). The recovery time, $t_R$, exhibits its first peak at 0.12 ms; this means that the longest $t_R$ is necessary when the initial polarized state is inverted by the $\pi$-pulse. We repeated a similar measurement, changing the peak-to-peak amplitude of the pulse, and found that the Rabi frequency indeed changes linearly, as shown in the right panel of Fig. \ref{rabi}(a). The characteristic decay time of the Rabi oscillation is $\sim $ 0.7 ms. This value is close to the coherence time of the nuclear spin, which is limited by a dipole-dipole interaction between neighboring nuclei \cite{taylor, yusa}. We also implemented the spin-echo measurement with three RF pulses \cite{machida} and found that the coherence time was almost the same as the decay time. Figure \ref{rabi}(b) shows the result of a pulsed $B_{rf}$ measurement with the $ST_+ -ST_+$ sequence. A similar sinusoidal oscillation was observed. Here, with the $\pi$-pulse of 0.12 ms, the initially antipolarized nuclear spins are inverted to the polarized direction, causing the maximum $t_R$ to return to the antipolarized state again.

Figure \ref{rabi}(c) shows the result of using the $ST_- -ST_+$ sequence, \textit{i.e.}, nuclear spins are initialized to the parallel polarization and rotated by a pulsed $B_{rf}$, and then the time required to switch to the antiparallel polarization is measured as $t_R$. The observed Rabi oscillation shows a phase shift of magnitude $\pi$, or equivalently, the Rabi oscillation is inverted. Note that parameters such as the amplitude and frequency of $B_{rf}$ and $B_{dc}$ are the same as those in the cases of Fig. \ref{rabi}(a) and (b). Only $V_{sd}$ is different. The recovery time, $t_R$, reaches its minimum value at the same $\pi$-pulse of 0.12 ms. This means that, after the $\pi$-pulse, polarized nuclear spins are inverted and become antiparallel, and this is identical to the nuclear spin state obtained at $V_{ST_+}$. Figure \ref{rabi}(d) shows the result for the $ST_+ -ST_-$ sequence; the $\pi$-shifted Rabi oscillation is again observed. These results clearly show that the parallel polarization obtained at $V_{ST_-}$ and the antiparallel polarization at $V_{ST_+}$ are linked by the NMR $\pi$-pulse.

In summary, it was demonstrated that nuclear spins in quantum dots can be polarized, and their direction can be switched to either parallel or antiparallel to the external magnetic field simply by switching the dc voltage applied to the device. With the use of the NMR $\pi$-pulse, it was confirmed that these polarization directions are indeed opposite each other. These results agree with the expected nuclear polarization mechanism of voltage-selective spin scattering from $T_-$ or $T_+$ to the singlet. From Rabi oscillation measurements, the coherent time of the polarized/antipolarized nuclear spins was found to be $\sim $ 0.7 ms. Decoupling pulse sequence may further enhance the coherence time as successively demonstrated for electron spin \cite{DasSarma, hanson2}. The results suggest that nuclear spins in quantum dot devices are an electrically accessible resource for future quantum information processing, such as a long-lived quantum memory for electron spin qubits.

We thank H. Akimoto, M. Kawamura and S. Amaha for useful discussions and assistance with the experiments. This work was supported by CREST-JST, ICORP-JST and a RIKEN Junior Research Associate (JRA) scholarship.


\newpage 

\end{document}